\documentstyle[12pt]{article}

\def\boxit#1{ \vbox{\hrule
height0.5pt\hbox{\vrule width0.5pt\kern10pt\vbox{ \kern10pt#1\kern10pt}
\kern10pt\vrule width0.5pt}\hrule height0.5pt}}

\def\bild#1\over#2{\mathrel{\mathop{\kern5pt #1}\limits_{#2}}}

\title{Tunneling and the Band Structure of Chaotic Systems}
\author{}
\date{{\Large P. Leb{\oe}uf and A. Mouchet}\\Division de Physique
Th\'eorique\footnote{Unit\'e de recherche des Universit\'es de Paris XI et
Paris VI associ\'ee au CNRS}\\ Institut de Physique Nucl\'eaire\\91406 Orsay
Cedex, France \vspace{0.08in} \\}

\begin{document}

\baselineskip 0.35in
\setcounter{page}{1}
\maketitle

\vspace{0.8in}

\begin{abstract}
We compute the dispersion laws of chaotic periodic systems using the
semiclassical periodic orbit theory to approximate the trace of the powers of
the evolution operator. Aside from the usual real trajectories, we also include
complex orbits. These turn out to be fundamental for a proper description
of the band structure since they incorporate conduction processes through
tunneling mechanisms. The results obtained, illustrated with the kicked-Harper
model, are in excellent agreement with numerical simulations, even in the
extreme quantum regime.
\end{abstract}
\vspace{0.2in}

\noindent IPNO/TH 94-19 ~~~~~~~~~~~~~~~~~~~~~~~~~~~~~~~~~~~~~~~~~~~~~~~~~~~~~
{}~~~~~~~~~~~~~~~~~~~~~~~~~~~~~ Avril 1994\\
%\noindent PACS numbers: 05.45.+b; 71.25.Cx; 03.65.Sq

\pagebreak

As is well known, the quantum motion of electrons in periodic potentials gives
rise  to energy bands. A precise determination of the density of states is
necessary in order to obtain an appropriate description of the elementary
physical properties, like the transport and magnetic properties. However, a
direct calculation of the dispersion laws is in general difficult, in
particular if the underlying classical dynamics is chaotic.

The purpose of this letter is to consider the band structure of classically
chaotic periodic systems from the point of view of periodic orbit theory. In
this approach, the trace of the evolution operator -- which can be related to
the density of states -- is expressed, in the semiclassical limit, as a sum
over classical periodic paths \cite{gutz,bb1}. As shown below, the
dependence of the dispersion laws on the quasimomenta (or Bloch angles) is
related to the winding numbers of the periodic orbits (p.o.'s) around the
elementary cell. Trajectories with winding numbers different from zero are
associated to open diffusive processes through the periodic lattice. However,
in some cases these  processes are classically forbidden, or not well described
by the real p.o.'s. The relevant features of the band structure can however be
recovered going one step beyond this approximation by including some tunneling
effects in the semiclassical trace formula. This is done by taking into account
the {\sl complex} p.o.'s (by complex or ghost p.o.'s we mean the periodic
solutions of Hamilton's equations having at least one complex phase-space
coordinate). Moreover, the inclusion of ghost orbits extends the range of
validity of the semiclassical approximation. As an illustration, at the end of
the Letter we will present an example in which, in spite of being far from a
classical behaviour, we are able to compute the (quasi-)energies as a function
of the Bloch-numbers (and hence the transport properties of the system) in
terms of complex p.o.'s, while only taking into account the real p.o.'s gives a
very crude approximation to the true solution.

In the standard WKB theory -- valid for integrable systems -- complex
classical paths have been used in the past in different contexts, as in the
study of the band structure of one-dimensional periodic potentials. To our
knowledge, the existence and relevance of ghost orbits in the semiclassical
trace formula was pointed out for the first time in \cite{bb2,bb1}. More
recently, these orbits were studied in more detail in Refs.\cite{khd,ss} in
connection to the trace of the evolution operator of the kicked top and the
standard map, respectively.  Here, we also consider a dynamics described by a
discrete (chaotic) map, inspired by the study of the motion of an electron in a
two-dimensional periodic potential in the presence of a uniform and constant
magnetic field perpendicular to the crystal plane. For a strong magnetic field,
the lowest Landau level approximation leads to the motion of a $1$D-particle
whose phase space is a $2$D-torus ${\cal T}$. If Planck's constant is rational,
$h=M/N$, then the corresponding evolution operator $U$ is doubly periodic and
its spectrum is made of $N$ bands $\epsilon_\alpha ({\vec{\, \theta}})$
given by $U |\psi_\alpha (\vec{\, \theta}) > = \exp({\rm i}\epsilon_\alpha
({\vec{\, \theta}})) |\psi_\alpha (\vec{\, \theta}) >$, $\alpha=1,\ldots,N$.
For simplicity, in the following we take $M=1$ and ${\cal T}$ will be of unit
length in both directions. The Bloch parameters $\vec{\, \theta}=\left(
\theta_1,\theta_2\right)   $ are associated to generalized periodic boundary
conditions $\psi(q+1) = \nolinebreak \exp\left( {\rm i}\theta_1\right)
\psi(q), \;\psi(p-1) = \exp\left( {\rm i}\theta_2\right) \psi(p)$. The
classical limit is obtained when $N\to\infty$.

In order to mimic the motion of the electron as the magnetic field is lowered,
we consider the kicked Hamiltonian $H=f(p)+g(q)\sum\limits_{n=-\infty}^{\infty}
\delta(t-n\tau)$, which leads to a discrete map on the torus
\begin{equation}\label{1}
\left\{ \begin{array}{lllr} q_{i+1} & = &q_i+\tau f'(p_{i+1}) & \;\; {\rm mod}
1
 \\ p_{i+1} & = &p_i-\tau g'(q_i) & \;\;\;\;\;\; {\rm mod} 1,
\end{array} \right.
\end{equation}
where $\left(   q_j,p_j\right)    \in [0,1[^2$ are the phase-space coordinates
of the particle at time $t=j\tau$ while $f$ and $g$ are two smooth functions of
period one. One can interpret the two integers
$$w_q^{(i+1)}=\left[ q_i+\tau f'(p_{i+1})\right]  ,\;\; w_p^{(i)}=-\left[
p_i-\tau g'(q_i)\right]  $$
(square brackets denote integer part) as the winding numbers of the orbit
around the torus for the iteration $\left( q_i,p_i\right) \to \left( q_{i+1},
p_{i+1} \right)$ in the $q$ and $p$ directions, respectively. As we shall see
in the following these two integers -- which arise from the toroidal topology
of phase space by the identification of points differing by an integral number
of phase-space cells -- play a crucial role in the semiclassical interpretation
of the bands (see also Ref.\cite{robbins}). Quantum mechanically the map is
implemented by the one-step evolution operator
\begin{equation} \label{2}
U={\rm e}^{-i\tau f(\hat{p})/\hbar} {\rm e}^{-i\tau g(\hat{q})/\hbar}.
\end{equation}
In order to make explicit the dependence on the parameters $\vec{\, \theta}$,
instead of $U$ we consider the operator ${\tilde U} (\vec{\, \theta})$ which is
obtained from Eq.(\ref{2}) by making the substitutions
$\hat{q}\to\hat{q}+\hbar\theta_2,\linebreak \hat{p}\to\hat{p}+ \hbar\theta_1$
\cite{lkfa}. Then the dispersion laws $\epsilon_\alpha ({\vec{\, \theta}})$ are
given by the roots $z_\alpha(\vec{\, \theta})=\exp({\rm i}\epsilon_\alpha
({\vec{\, \theta}})) \linebreak\alpha=1,\ldots,N$ of the spectral determinant
\begin{equation} \label{3}
\det\left( z-{\tilde U} (\vec{\, \theta})\right) =\sum_{k=0}^N
 a_k(\vec{\, \theta})z^k,
\end{equation}
where $a_N(\vec{\, \theta})=1$ by construction. The coefficients of the
characteristic polynomial (\ref{3}) can be obtained from the trace of the first
$N$ powers of ${\tilde U}$ through the formula
\begin{equation} \label{4}
a_{N-k}=-\frac{1}{k}\sum_{n=1}^k a_{N-k+n}{\mbox{\rm tr}}\,({\tilde U}^n),
\;\;\;\;\;\; k=1,\ldots,N.
\end{equation}

The interest of (\ref{4}) comes from the fact that in the semiclassical limit
$N\to\infty$ the trace of the powers of ${\tilde U}$ can be computed in terms
of a finite number of  classical p.o.'s (which are assumed to be isolated)
\cite{gutz,tabor}. In our case, by considering the special topology of phase
space, we get
\begin{equation} \label{5}
{\mbox{\rm tr}}\,({\tilde U}^n)={\rm i}^n\sum_{\rm p.o.}
\frac{T}{\sqrt{|\det(M-1)|}} \exp\left(   \frac{{\rm i}}{\hbar}S +{\rm i}
w_p\theta_2+{\rm i} w_q\theta_1-{\rm i}\frac{\pi}{2}\nu\right)   .
\end{equation}
In this formula the sum extends over all the real p.o.'s of the map (\ref{1})
whose period $T$ is an integer fraction of $n$, $w_p$ and $w_q$ are the total
winding numbers of the p.o., $w_p=\sum\limits_{j=0}^{n-1} w_p^{(j)}, \, w_q =
\sum\limits_{j=0}^{n-1} w_q^{(j)}$,\linebreak $S$ is the classical action of
the p.o.
$$S=\sum_{j=0}^{n-1}\left\{-\tau\left( g(q_j)+f(p_j)\right)
+p_{j+1}(q_{j+1}-q_j) - w_p^{(j)}q_j-w_q^{(j)}p_j\right\},$$
$$M=\prod_{j=n}^1\left(\begin{array}{cc} 1-\tau^2g''(q_{j-1})f''(p_j)&\tau
f''(p_j)\\ -\tau g''(q_{j-1})&1 \end{array}\right)$$
is the monodromy matrix (related to its stability) and $\nu$ is the number of
negative eigenvalues of the $2n\times 2n$ matrix (assumed to be non singular):
$$\left(\begin{array}{cc} \left(   \frac{\partial^2S}{\partial p_i\partial
p_j}\right)   _{ i,j=1,\ldots,n}&\left(   \frac{\partial^2S}{\partial
p_i\partial q_j}\right)   _{i,j=1,\ldots,n} \\ \left(
\frac{\partial^2S}{\partial q_i\partial p_j}\right)   _ {i,j=1,\ldots,n}&
\left(   \frac{\partial^2S}{\partial q_i\partial q_j}\right)   _
{i,j=1,\ldots,n} \end{array}\right). $$
The information concerning the band structure (i.e., the dependence on $\vec{\,
\theta}$ of the dispersion laws) is thus recovered, semiclassically, through
the factor $\exp\left( {\rm i} \left( w_q \theta_1+ w_p\theta_2\right)
\right)$.  This factor, which is different from zero for open diffusive
p.o.'s, can be interpreted for each $\vec{\, \theta}$ as a unitary
representation of the homotopy group of ${\cal T}$. The fact that the
representation of that group appears in ${\mbox{\rm tr}}\,( {\tilde U}^n)$ when
one deals with path integrals (or their discrete equivalent) in
multiply-connected spaces is well known \cite{schul}.

Because $U$ is a unitary operator, the coefficients in Eq.(\ref{3}) satisfy the
self-inversive property \cite{marden}
\begin{equation} \label{6}
a_{N-k}=\exp\left({\rm i}\Theta\right) \overline{a_k}\, , \;\;\; k=0,\ldots,N
\end{equation}
where $\Theta=\pi N+\sum \epsilon_\alpha ({\vec{\, \theta}})$. This property,
together with Eqs.(\ref{3})-(\ref{5}), allows to compute the band spectrum in
terms of the classical p.o.'s of length less than $[N/2]$.

Eq.(\ref{5}) was obtained from the exact expression of ${\mbox{\rm
tr}}\,({\tilde U}^n)$ replacing the sums by integrals using the Poisson
summation formula and then computing the integrals by the stationary phase
approximation. One step further is to deform the contour of integration of the
path integral in order to reach the stationary orbits lying in the complex
plane. The outcome of this procedure is to include in the summation (\ref{5})
complex p.o.'s, whose contribution is exponentially small. However, as we shall
now see, they cannot be neglected if $\hbar$ is not too small or if we are
close to a (first order) bifurcation of a p.o. .

For the sake of definiteness, we consider the kicked-Harper model
\begin{equation}
\label{7} g(x)=f(x)=-\gamma\cos(2\pi x)/(2 \pi \tau).
\end{equation}
Phase-space plots of the classical trajectories of the kicked-Harper map for
different values of $\gamma>0$ can be found in \cite{lkfa}: for $\gamma \to
0^+$ the dynamics is integrable (it tends to the Harper Hamiltonian); for
$\gamma\simeq 0.4$ it is a mixture of regular and chaotic orbits, while for
$\gamma \geq 0.6$ unstable chaotic trajectories dominate. At the quantum level,
some spectral and transport properties of the model were studied in
\cite{lkfa,kh}.

We now compute the trace of ${\tilde U}$ for this model. The exact result is
\begin{equation}
\label{8} {\mbox{\rm tr}}\, {\tilde U}=N\left\{ J_0(\gamma
N)+2\sum_{w=1}^\infty{\rm e}^ {-{\rm i} wN\pi/2}\cos(w\theta_1)J_{wN}(\gamma
N)\right\}   \times  \left\{ \theta_1\rightleftharpoons \theta_2\right\}  ,
\end{equation}
where the second curly brackets in the r.h.s. is identical to the first one
except for the fact that $\theta_1$ is replaced by $\theta_2$. In order to
compute (\ref{8}) semiclassically, we need to find all the p.o.'s of period one
of the map. Their location $(q,p)\in[0,1[^2$ is given, from (\ref{1}) and
(\ref{7}), by the solutions of the set of equations
\begin{equation} \label{9}
\left\{\begin{array}{ccc} \sin (2\pi q)&=& w_p/\gamma\\ \sin (2\pi p)&=&
w_q/\gamma, \end{array} \right.
\end{equation}
where $(w_q,w_p)\in{\mbox{Z$\!\!$Z}}^2$. Let us briefly discuss the different
solutions of Eq.(\ref{9}). Assume for the moment that both $w_q$ and $w_p$ are
positive. A set of winding numbers $(w_q,w_p)$ corresponds to a {\sl real} p.o.
if $w_{q,p}\leq\gamma$. In that case there are four different solutions of
Eq.(\ref{9}) in $[0,1[^2$: $(q_0, p_0)$, $(q_0, 1/2-p_0)$, $(1/2-q_0, p_0)$ and
$(1/2-q_0, 1/2-p_0)$, where $q_0=\arcsin(w_p/\gamma)/2\pi$ and $p_0=
\arcsin(w_q/\gamma)/2\pi$. On the other hand, $(w_q,w_p)$ correspond to a {\sl
complex} p.o. if at least one of its winding numbers is larger than $\gamma$.
If both integers are larger than $\gamma$, there are four different p.o.'s
located at
\begin{equation} \label{10}
\left\{\begin{array}{ccc} q&=&1/4\pm{\rm i}{\mbox{\rm
Arch}}\,(w_q/\gamma)/2\pi\\ p&=&1/4\pm{\rm i}{\mbox{\rm
Arch}}\,(w_p/\gamma)/2\pi; \end{array} \right.
\end{equation}
while if $w_i<\gamma<w_j$ one of the coordinates is real and the other
one is complex.

For negative values of $w_q$ and/or $w_p$, the same kind of analysis can be
done. The global description of the period-one p.o.'s is therefore as follows.
For a fixed $\gamma>0$, there is an infinite number of complex p.o.'s (all
integers $(w_q,w_p)$ whose modulus is larger than $\gamma$). Since the
imaginary part of the coordinate is proportional to ${\mbox{\rm Arch}}\,
|w_i/\gamma|$, the complex p.o.'s with large winding number $|w_i|>>\gamma$ are
deeply located in the complex plane.  As $\gamma$ increases (starting from
$1/2$, for example), a complex orbit becomes real if $\gamma$ becomes larger
than the modulus of both winding numbers.  This will happen at each integer
value of $\gamma$ through a first-order bifurcation (fold catastrophe), where
several orbits coalesce in at least one phase-space coordinate. The total
number of real p.o.'s increases with $\gamma$ as $(4 [\gamma]+2)^2$.

Concerning the stability of those orbits, generically ${\mbox{\rm tr}}\, M=2$
at $\gamma_n=n\in{\mbox{I$\!$N}}^*$ for all the real p.o.'s emerging at that
value of $\gamma$. When $\gamma$ is increased, half of them are initially
stable, the other half unstable; but very rapidly they become all unstable
\cite{fn1}.

Using the information described above we have computed semi-classically the
trace of ${\tilde U}$ (for fixed $\gamma$ and arbitrary $N$) including all the
real as well as complex p.o.'s. The result is
\begin{equation} \label{11}
{\mbox{\rm tr}}\, {\tilde U} = \frac{2}{\pi\gamma} F(\theta_1)
F(\theta_2)+{\cal O}(1/N)
\end{equation}
where
\begin{eqnarray} \label{12}
F(\theta) & = & 2 \sum_{w=0}^{[\gamma]}{}\!' {\rm e}^{-{\rm i} w N\pi/2}
\frac{\cos(w\theta)}{\sqrt{\cos(x_w)}} \cos \left\{ N\left( \gamma\cos(x_w)- w
(x_w-\pi/2)\right)  -\pi/4\right\} \nonumber \\ & + &\sum_{w=[\gamma]+1}^\infty
{\rm e}^{{\rm i} w N\pi/2} \frac{\cos(w\theta)} {\sqrt{{\mbox{\rm
sinh}}\,(y_w)}} {\rm e}^{-N \left( w y_w-\gamma{\rm sinh}\,(y_w)\right)}
\end{eqnarray}
and
\begin{equation}
\left\{\begin{array}{ccc} x_w&=&\arcsin(w/\gamma) \nonumber  \\
y_w&=&{\mbox{\rm Arch}}\,(w/\gamma). \nonumber \end{array} \right.
\end{equation}
In Eq.(\ref{11}), $F(\theta_2)$ is related to the $q$-coordinate of the orbit,
while $F(\theta_1)$ contains the information about its momentum. The first sum
in Eq.(\ref{12}) ($w\leq[\gamma]$) comes from p.o.'s having a real coordinate
(which reproduces the oscillating part of Bessel functions occurring in
Eq.(\ref{8})), while the second sum ($w>[\gamma]$) comes from those having a
complex coordinate (reproducing the exponentially decreasing part of the Bessel
functions) . The symbol $\sum^{'}$ indicates that the term $w=0$ must be
multiplied by a factor $1/2$. The product of both functions in Eq.(\ref{11})
gives rise to four different sums, one related to real p.o.'s, the other three
to complex ones. The dependence on $\vec{\, \theta}$ in Eq.(\ref{11}) is the
same for both kinds of orbits, but the complex orbits are exponentially dumped
by a factor (or a product of factors) of the form $\exp(-N S_I)$, where $S_I =
w y_w-\gamma {\mbox{\rm sinh}}\,(y_w)>0$. Because of that, complex p.o.'s with
large winding number can in general be neglected. On the other hand, as
$\gamma$ approaches some integer value $n$ from below, the contribution of the
complex p.o.'s becoming real at $\gamma_n=n>0$ is particularly important,
since the imaginary part of their action tends to zero. In fact, they remain
important for parameter values relatively far from $\gamma_n$, since $S_I\simeq
\sqrt{2/n} (\gamma_n - \gamma)^{3/2}/3\pi$ as $\gamma\to\gamma_n$ from below
(the exponent $3/2$ was also found for the kicked top and the standard map
\cite{khd,ss}). Moreover, because of the $1/\sqrt{n}$ dependence, the parameter
interval in which complex orbits are important increases with $\gamma$ (at a
fix $N$). Exactly at $\gamma=\gamma_n$ Eq.(\ref{11}) is not defined since the
denominator of both sums in $F(\theta)$ vanishes due to the coalescence of
several orbits. In order to avoid divergences we must improve the
approximation, and include third order terms in the computation of the
integrals for $\gamma\simeq \gamma_n$ using Airy functions \cite{bm,bb2}.
Including these corrections, we find that (\ref{11}) reproduces extremely well
the exact ${\mbox{\rm tr}}\, {\tilde U}$, even for small values of $N$.

In order to illustrate this, let us consider the {\sl extreme quantum limit}
$N=2$. For the kicked-Harper map, it can be shown that $\det {\tilde U}=1$ for
arbitrary $\gamma$, $N$ and $\vec{\, \theta}$, and then $\Theta=\pi N$ in
Eq.(\ref{6}). From this it follows that for $N=2$ all the information
concerning the spectrum of ${\tilde U}$ is contained in ${\mbox{\rm tr}}\,
{\tilde U}$, since Eq.(\ref{3}) reduces to a second-degree polynomial with
coefficient $a_2=1, a_1=-{\mbox{\rm tr}}\, {\tilde U}, a_0=1$. The two
dispersion relations obtained from that equation are $\epsilon_{\pm}(\vec{\,
\theta})=\pm \arccos({\mbox{\rm tr}}\, {\tilde U}(\vec{\, \theta})/2)$.
Although our results are valid for arbitrary $\gamma>0$, the role of the
complex p.o.'s is particularly stressed in $0<\gamma<1$. In that parameter
interval there exist only four real p.o.'s of period one, located at $(0,0)$,
$(0,1/2)$, $(1/2,0)$, $(1/2,1/2)$. But these orbits do not introduce any
$\vec{\, \theta}$ dependence in ${\mbox{\rm tr}}\, {\tilde U}$ (and therefore
in $\epsilon_\pm (\vec{\, \theta})$) since for all of them $w_q=w_p=0$, leading
to flat bands. In fact, the exact dispersion laws for $N=2$ have a strong and
non-trivial $\vec{\, \theta}$-dependence; in particular, for an arbitrary
$\gamma$ there is at least one linear contact (diabolical point) between the
two bands. In order to introduce semiclassically a $\vec{\, \theta}$ dependence
in the dispersion laws one must include complex p.o.'s. Fig.1(a) plots the
exact result and the semiclassical dispersion laws computed from Eq.(\ref{11})
with and without complex orbits for $\gamma=0.86$ and $\theta_2/2\pi=0.5$ as a
function of $\theta_1/2\pi$ (only complex p.o.'s up to winding number two were
included). The result obtained with complex p.o.'s is extremely good even near
the diabolical point, which is the worst situation for semiclassical analysis.
Part (b) of that Figure shows the case $\gamma=4.86$.

It is also possible to compute, using only the ${\mbox{\rm tr}}\, {\tilde U}$,
the dispersion laws for $N=3$ since by the self-inversive symmetry and the fact
that $\Theta=3\pi$ the coefficients of the characteristic polynomial are now
given by  $a_3=1,\; a_2=-{\mbox{\rm tr}}\, {\tilde U},\;
a_1=\overline{{\mbox{\rm tr}}\, {\tilde U}},\; a_0=1$. Fig.1(c)-(f) illustrate
the results obtained in this case for the dispersion laws and the trace of
${\tilde U}$.

As a final remark, let us point out that the approach presented in this Letter
is also relevant in systems where the topology producing the band structure
is not toroidal, but for example cylindrical. This occurs in particular in the
context of persistent currents, where the Aharonov-Bohm flux threading the ring
plays the role of the Bloch-parameter and the winding number of the p.o.'s is
precisely the winding number around that flux (see Ref.\cite{moriond}
for a more detailed discussion).

\pagebreak

\pagebreak
\large
\begin{center}
FIGURE CAPTION
\end{center}
\normalsize

\begin{description}

\item{FIG. 1:} Band structure and trace of the evolution operator for the
kicked-Harper model. (a) dispersion laws $\epsilon (\vec{\, \theta})$ as a
function of $\theta_1/2\pi$ for $\theta_2/2\pi=0.5$, $N=2$ and $\gamma=0.86$.
Full line: exact; dotted line: semiclassical theory including the real and
complex p.o.'s; dashed line: semiclassical theory including only the real
p.o.'s.
(b) same as in part (a) but for $\gamma=4.86$ and $\theta_2/2\pi=0.25$. (c)-(d)
same as in part (a) but for $N=3$, $\theta_2/2\pi=0.25$, $\gamma=2.86$ and
$\gamma=52.1$, respectively. (e)-(f) real and imaginary part of ${\mbox{\rm
tr}}\,({\tilde U})$ as a function of $\gamma$ for $\theta_1/2\pi=0.55$,
$\theta_2/2\pi=0.32$ and $N=3$. The semiclassical result using only the real
p.o.'s is not shown in (e)-(f). Except near the diabolical point in part (a),
in all the other plots the dotted line cannot in general be distinguished from
the full line.
\end{description}

\baselineskip 0.26in
\begin{thebibliography}{99}

\bibitem{gutz} M. C. Gutzwiller, {\sl J. Math. Phys.} {\bf 12}, 343 (1971).
\bibitem{bb1} R. Balian and C. Bloch, {\sl Ann. Phys.} (NY) {\bf 85}, 514
 (1974).
\bibitem{bb2} R. Balian and C. Bloch, {\sl Ann. Phys.} (NY) {\bf 69},
 76 (1972).
\bibitem{khd} M. Ku\'s, F. Haake and D. Delande, {\sl Phys. Rev. Lett.} {\bf
 71}, 2167 (1993).
\bibitem{ss} R. Scharf and B. Sundaram, preprint.
\bibitem{robbins} J. M. Robbins, {\sl Phys. Rev. A} {\bf 40}, 2128 (1989).
\bibitem{lkfa} P. Leb{\oe}uf, J. Kurchan, M. Feingold and D.P. Arovas,  {\sl
 Phys. Rev. Lett.} {\bf 65}, 3076 (1990); CHAOS {\bf 2}, 125 (1992).
\bibitem{tabor} M. Tabor {\sl Physica} {\bf 6D}, 195 (1983); G. Junker and
H. Leschke {\sl Physica} {\bf D 56}, 135 (1992).
\bibitem{schul} L. S. Schulman {\sl Techniques and Applications of Path
Integration} (Wiley, New York, 1981),  chap. 23.
\bibitem{marden} M. Marden, {\sl Geometry of Polynomials} (Am. Math. Soc.,
Providence, 1966).
\bibitem{kh} R. Lima and D.L. Shepelyansky, {\sl Phys. Rev. Lett.} {\bf 67},
1377 (1991); T. Geisel et al, {\sl Phys. Rev. Lett.} {\bf 67}, 3635 (1991);
R. Artuso, G. Casati and D. Shepelyansky, {\sl Phys. Rev. Lett.} {\bf 68}, 3826
(1992); R. Artuso et al, {\sl Phys. Rev. Lett.} {\bf 69}, 3302 (1992).
\bibitem{fn1} However, the existence of a small parameter range (whose size
decreases with $\gamma$) above each integer value of $\gamma$ around which some
of the period-one p.o.'s are stable demonstrates that the kicked-Harper map,
even for large values of $\gamma$, is never completely hyperbolic.
\bibitem{bm} M. V. Berry and K. E. Mount {\sl Rep. Prog. Phys.} {\bf 35}
315 (1972); A. Ozorio de Almeida and J. Hannay, J. Phys. A {\bf 20}, 5873
(1987).
\bibitem{moriond} P. Leb{\oe}uf and A. Mouchet, in preparation.
\end{thebibliography}
\end{document}